\documentclass[submission, Phys]{SciPost}
\hypersetup{
    colorlinks,
    linkcolor={red!50!black},
    citecolor={blue!50!black},
    urlcolor={blue!80!black}
}

\usepackage[bitstream-charter]{mathdesign}
\DeclareSymbolFont{usualmathcal}{OMS}{cmsy}{m}{n}
\DeclareSymbolFontAlphabet{\mathcal}{usualmathcal}

\usepackage{subfig}
\usepackage{bbm}
\usepackage[normalem]{ulem}

\newcommand{\odiff}[2]{\frac{{\rm d}#1}{{\rm d}#2}}

\newcommand{\Tr}{{\rm Tr}}

\def\nn{\nonumber \\ }

\newcommand{\R}{\mathbbm{R}}
\newcommand{\Z}{\mathbbm{Z}}
\newcommand{\cop}[1]{ c_{#1}^{\phantom{\dag}} }

\renewcommand{\Re}{\operatorname{Re}}
\renewcommand{\Im}{\operatorname{Im}}

\DeclareMathOperator{\Pf}{Pf}

\newcommand{\be}{\begin{equation}}
\newcommand{\ee}{\end{equation}}
\newcommand{\bea}{\begin{eqnarray}}
\newcommand{\eea}{\end{eqnarray}}

\def\nn{\nonumber\\}
\def\fr#1{(\ref{#1})}
\begin{document}

\begin{center}{\Large \textbf{
      A simple theory for quantum quenches in the ANNNI model}}
\end{center}

\begin{center}
    Jacob H. Robertson\textsuperscript{1$\star$},
    Riccardo Senese\textsuperscript{1} and
    Fabian H. L. Essler\textsuperscript{1}
\end{center}

\begin{center}
{\bf 1} The Rudolf Peierls Centre for Theoretical Physics, Oxford University, Oxford OX1 3NP, UK
\\
${}^\star$ {\small \sf  jacob.robertson@physics.ox.ac.uk}
\end{center}

\begin{center}
\end{center}

\section*{Abstract}
{\bf
In a recent numerical studyby Haldar et al. (Phys. Rev. X \textbf{11}, 031062) it was shown that
signatures of proximate quantum critical points can be observed at
early and intermediate times after
certain quantum quenches. Said work focused mainly on the case of the
axial next-nearest neighbour Ising (ANNNI) model. Here we
construct a simple time-dependent mean-field theory that allows us to 
obtain a quantitatively accurate description of these quenches at
short times, which for reasons we explain remains a fair approximation
at late times (with some caveats). Our approach provides a simple
framework for understanding the reported numerical results as well as
fundamental limitations on detecting quantum critical points through
quench dynamics. We moreover explain the origin of the peculiar
oscillatory behaviour seen in various observables as arising from the
formation of a long-lived bound state. \phantom{\cite{Haldar_2021}} 
}

\vspace{10pt}
\noindent\rule{\textwidth}{1pt}
\tableofcontents\thispagestyle{fancy}
\noindent\rule{\textwidth}{1pt}
\vspace{10pt}

\section{Introduction}
\label{sec:intro}

Quantum phase transitions (QPT) provide a key framework for our
understanding of equilibrium phases of correlated quantum matter
\cite{sachdev2000quantum}. More recently physical properties in the
vicinity of quantum critical points in out-of-equilibrium settings
have been investigated theoretically
\cite{calabrese2016quantum,EsslerQuench2016,cazalilla2016quantum} 
and in ultra-cold atom experiments
\cite{Langen2015Ultracold,Zeiher2017Coherent,bouchoule2021generalized,malvania2021generalized}. An
interesting question that has been raised is whether it is
possible to detect the location of QPTs, and associated physical
properties, through the dynamics at short and intermediate times after
a quantum quench
\cite{Bhattacharyya2015Signature,Heyl2018Detecting,Titum2019Probing,Dag2021Detecting,Haldar_2021,Paul2022Hidden}.
In Ref. \cite{Haldar_2021} Haldar et al. proposed a set of generalized
susceptibilities that quantify the sensitivity of the time evolution
and stationary values of local observables to changes in the quench
protocol. Based on numerical studies in the axial next-nearest neighbour Ising model
(ANNNI) the authors concluded that such susceptibilities can
indeed provide signatures of a proximate QPT not only in the
stationary regime but already at short/intermediate times. An
important question is how general this approach is, and what its
limitations are. In order to address these issues we show that
the findings of Ref.~\cite{Haldar_2021} for the ANNNI model can be
understood in terms of a simple (time-dependent) mean-field
theory. This approach provides a clear insight into the window
of applicability of any approach using generalized susceptibilities to
search for the location of critical points. En route we clarify the
origin of interesting oscillatory behaviours of local observables
observed in Ref.~\cite{Haldar_2021}.

The outline of this paper is as follows. In Section \ref{Sec:Model} we
introduce the ANNNI model and describe the quench protocol we
consider. In Section \ref{Sec:SCMFT} we then construct a mean-field
description of the stationary state under the assumption that the
system thermalizes. In Section \ref{Sec:TDMFT} we construct a
time-dependent 
self-consistent mean-field description of the time evolution. Within 
this approximation the density matrix is Gaussian at all times and
Wick's theorem may be employed to calculate any correlation
function. This method is expected to be quantitatively accurate for 
short times as long as the initial state is itself Gaussian. In
Section \ref{Sec:NonEqualTimes} we show that non-equal time correlation
functions are easily accessible with this method and use it to compute
the transverse component of the generalized dynamical structure factor following a
quench in the ANNNI, demonstrating that this object contains
information about the spectrum of the post-quench Hamiltonian.

\section{Definition of the model and quench protocol}
\label{Sec:Model}

The ANNNI model is a well studied non-integrable model with competing
interactions, see e.g. \cite{Peschel1981Calculation, Selke1988ANNNI,Chandra2007Floating,
  Karrasch2013Dynamical}. The model consists of a transverse-field
Ising model with an additional next-nearest neighbour Ising exchange,
which we take to have the opposite sign to the nearest-neighbour Ising
interaction
\begin{equation}
H(h,\kappa) = -J \sum_{i}^{L} \sigma^x_i \sigma^x_{i+1}-h\sum_i \sigma_i^z + \kappa \sum_i^L \sigma_i^x \sigma^x_{i+2} \label{Eq:ANNNI_HSpins} \ . 
\end{equation}
Here $\sigma_j^\alpha$ are the usual Pauli matrices on sites $j$ of a
ring of circumference $L$. The Hamiltonian \fr{Eq:ANNNI_HSpins}
can be mapped to a model of spinless lattice fermions by means of a
Jordan-Wigner transformation\cite{lieb1961two}. As we adopt periodic boundary conditions
for the spins the fermions must obey either anti-periodic 
(Neveu-Schwarz) or periodic (Ramond) boundary conditions depending on
whether the fermion number is even or odd, see
e.g. Appendix A of \cite{CEF1}. In what follows we only consider operators which have even fermion parity, therefore for our purposes it is sufficient to
work in the Neveu-Schwarz sector for even system sizes $L$. The
Hamiltonian then reads 
\begin{align}
H(h,\kappa) =& -J\sum_j \left(c_j^\dag \cop{j+1} + c^\dag_j c^\dag_{j+1}+{\rm
  h.c.}\right) + \kappa\sum_j(c_j^\dag \cop{j+2} + c^\dag_j 
 c^\dag_{j+2}+{\rm h.c.}) + 2h\sum_j c_j^\dag \cop{j} \nn
&+ 2\kappa \sum_j\left(\cop{j} c^\dag_{j+1} \cop{j+1} c^\dag_{j+2} - c_j^\dag c^\dag_{j+1} \cop{j+1} c^\dag_{j+2} + {\rm h.c.} \right) \label{Eq:ANNNI_HFerms}\ .
\end{align}
The next-nearest neighbour spin-spin interaction is seen to give rise to a
quartic interaction amongst the fermions, making the model non-integrable.
The Hamiltonian \fr{Eq:ANNNI_HSpins} has a global $\Z_2\otimes \Z_2$ symmetry
corresponding to rotations around the $z$-axis by $\pi$ -- which is
broken spontaneously in the ferromagnetic phase -- and site parity
$\sigma_n^\alpha \mapsto \sigma_{-n}^\alpha$. The latter remains
unbroken in the situations we consider and enforces
$t_{ij} \equiv \langle c_i^\dag \cop{j} \rangle =t_{ji}\in \R$ (see Appendix \ref{app:reality}), while the former
translates into fermion number parity.

\begin{figure}[ht]
   \centering
   \includegraphics{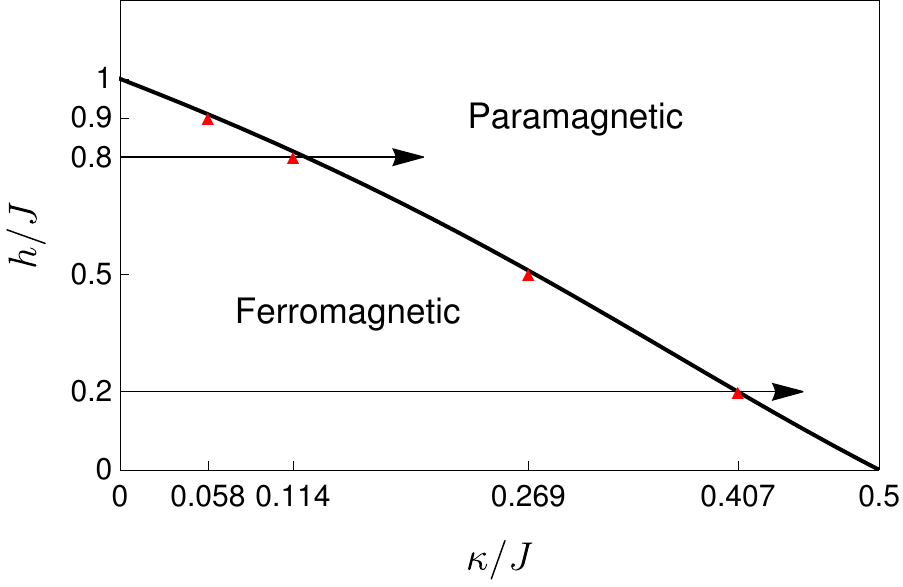}
   \caption{Ground state phase diagram of the ANNNI model for $0<\kappa/J<1/2$ - the solid curve is the boundary obtained by second order perturbation theory \fr{Eq:PhaseBdry2ndOrder}, red triangles indicate the critical points found by our self-consistent mean-field theory at select fields. There exist other phases at $\kappa>J/2$. Arrows indicate the quenches performed in this work.}
   \label{fig:PhaseDiagram}
\end{figure}

The ground state phase diagram of the ANNNI model for $\kappa<J/2$ is shown in
Fig.~\ref{fig:PhaseDiagram}
\cite{Selke1988ANNNI,allen2001two,Beccaria2007Evidence,suzuki2012quantum}. 
For $\kappa>J/2$ the phase diagram is substantially more complicated, but like Ref. \cite{Haldar_2021} we restrict our analysis to the ferromagnet-paramagnet transition.
At $\kappa=0$ the model \fr{Eq:ANNNI_HSpins} reduces to the transverse
field Ising model (TFIM) and is exactly solvable as it is quadratic in
fermions \cite{sachdev2000quantum,lieb1961two}. For $\kappa>0$ a second
order phase transition in the Ising universality class separates a
ferromagnetically ordered phase from a paramagnetic one. For $\kappa<J/2$ and small
values of $h$ the locus of the critical line can be determined by
second order perturbation theory, which yields \cite{Peschel1981Calculation} 
\begin{equation}
    J-2\kappa_c = h_c-\frac{1}{2J}\frac{\kappa_ch_c^2}{J-\kappa_c} \ . \label{Eq:PhaseBdry2ndOrder}
\end{equation}
In terms of the spins the transition is characterized by the
order parameter $\langle \sigma^x_j\rangle $ taking a non-zero value in the ferromagnetic phase. In terms of the fermions
this is a non-local (string) operator and the transition is topological
\cite{kitaev2001unpaired}. Our analysis of quench dynamics close to
quantum critical points in one dimension therefore pertains to both
topological transitions and conventional transitions with local order
parameters. Our mean-field analysis developed below may be expected to work rather well for small $\kappa/J$ as it is exact
along the line $\kappa=0$. Moreover, since the (equilibrium) mean-field theory we describe below is via a free fermion whose critical scaling limit is a relativistic Majorana fermion\cite{itzykson_drouffe_1989}, we correctly account for the symmetry and
critical exponents of the Ising transition. Hence our mean-field theory is
expected to give a quantitatively accurate description of the ANNNI
model in the region $h\approx J$ and $\kappa \approx 0$. Conversely, mean-field theory cannot be expected to give a reasonable description either deep into the paramagnetic phase or for the other transitions. We therefore do not investigate the rest of the phase diagram in this work.

In what follows we consider quantum quenches from initial thermal
states of the TFIM with transverse field $h_i$ and inverse temperature
$\beta$, i.e. our initial density matrix is
\be
\rho(t=0)=\frac{\exp\big(-\beta H(h_i , 0)\big)}{{\rm Tr} \exp\big(-\beta
  H(h_i , 0)\big)}\ .
\label{rhoinitial}
\ee
Including thermal states at finite temperatures rather than only ground states is useful as it allows us to tune the
energy density of the stationary state reached at late times in a
simple manner. We then consider the time evolution induced by the ANNNI
Hamiltonian $H(h_f,\kappa)$, i.e.
\be
\rho(t>0)=e^{-iH(h_f,\kappa)t}\rho(t=0)e^{iH(h_f,\kappa)t}\ .
\ee
We will restrict ourselves to the case $h_i=h_f\equiv h$ and quenches with
$\kappa<J/2$. To simplify notations we also set $J=1$. As the ANNNI
model is non-integrable when both $h$ and $\kappa$ are non-zero we
expect the model to thermalize
\cite{PolkovnokivNonequilibrium2011,EsslerQuench2016}, i.e. 
in the thermodynamic limit the system should locally relax to a thermal
stationary state described by an effective temperature that is set
by the energy density of the initial state
\be
e_0=\lim_{L\to\infty}\frac{1}{L}\text{Tr}\Big(\rho(t=0)H(h_f,\kappa)\Big). \label{Eq:InitialEnergy}
\ee
In our setup the correlation length typically starts off small
as a result of a large pre-quench gap, while at late times the system
settles into a thermal state at a low effective temperature in the vicinity of a
quantum critical point. Hence the correlation length in the stationary
state is typically much larger than in the initial state. Intuitively
therefore the physics should be that of a system whose correlation
length grows following the quench.

\section{Mean-field theory for the stationary state}
\label{Sec:SCMFT}
Since the ANNNI model is believed to thermalize and has no local conservation laws other than the total energy, we expect local
observables $\mathcal{O}$ to reach their Gibbs ensemble values at late
times after a quantum quench 
\begin{equation}
    \langle \mathcal{O} \rangle(t) \overset{t=\infty}{\longrightarrow} Z^{-1}\Tr[e^{-\beta_f H(h_f,\kappa)} \mathcal{O} ] \ \label{Eq:Thermal} .
\end{equation}
Here $Z$ is the partition function and $\beta_f$ the inverse effective temperature, set by the initial
energy density \fr{Eq:InitialEnergy} generated by the quench protocol.
For sufficiently small values of $\kappa$ this thermal state should be
amenable to a description in terms of a simple self-consistent
mean-field theory of spinless fermions
\begin{equation}
  Z^{-1}\Tr[e^{-\beta_f H} \mathcal{O} ]\approx
  Z_{\rm MFT}^{-1}\Tr[e^{-\beta_{\rm MFT} H_{\rm MFT}} \mathcal{O} ]\ ,
\end{equation}
where 
\begin{align}
H_{\rm MFT}= \sum_i \sum_{a=0}^2 \left\{J_{\rm Eff}^{(a)}(c^\dag_i c_{i+a} +{\rm hc})+ (\Delta_{\rm Eff}^{(a)} c^\dag_i c^\dag_{i+a}+{\rm hc}) \right\} +E_0 \ . \label{Eq:MFT_H}
\end{align}
This mean-field theory is the result of requiring that Wick's theorem holds, or equivalently that higher cumulants vanish. The effective couplings $J_{\rm Eff}^{(a)}$ and $\Delta_{\rm
  Eff}^{(a)}$ and the constant $E_0$ are generated by decoupling the quartic interaction terms
self-consistently via
\begin{equation}
    ABCD \mapsto \langle AB \rangle_{\rm MFT} CD + AB \langle CD \rangle_{\rm MFT} -
    \langle AB \rangle_{\rm MFT} \langle CD \rangle_{\rm MFT} + \text{all other Wick
      contractions}\ ,
\end{equation}
where
\be
\langle {\cal O}\rangle_{\rm MFT} \equiv
  Z_{\rm MFT}^{-1}\Tr[e^{-\beta_{\rm MFT} H_{\rm MFT}}{\cal O}]\ . \label{Eq:SelfConsistent}
\ee
Defining the (self-consistent) expectation values
\begin{align}
t_a& \equiv \langle c^\dagger_jc_{j+a}\rangle_{\rm MFT}\ ,\quad a=0,1,2\ ,\nn
\Delta_b& \equiv \langle c^\dagger_jc^\dagger_{j+b}\rangle_{\rm MFT} \ ,\quad b=1,2 \ , 
\end{align}
we have
\begin{align}
    \begin{array}{ll}
      J_{\rm Eff}^{(0)}= h-2\kappa (t_2+\Re \Delta_2)   \ ,&   \\[6pt]
      J_{\rm Eff}^{(1)}= -(J-4\kappa(t_1+\Re\Delta_1))  \ , & \Delta_{\rm Eff}^{(1)}= -(J -4 \kappa(t_1+\Delta_1^*)) \ , \\[6pt]
      J_{\rm Eff}^{(2)}= \kappa(1-2t_0) \ , & \Delta_{\rm Eff}^{(2)}= \kappa(1-2t_0) \ , \\[6pt]
       \multicolumn{2}{l}{E_0=-hL - 4L\kappa(|\Delta_1|^2+t_1^2-t_0t_2+2\Re\Delta_1 t_1 - \Re\Delta_2 t_0)} \ . 
    \end{array}
\label{Eq:Parameters}
\end{align}
Note that $J_{\rm Eff}^{(1)}\neq \Delta_{\rm Eff}^{(1)}$.
In order to fully specify our self-consistent mean-field theory we
require the self-consistent values of the five mean-fields as well as the
value of the inverse effective temperature $\beta_{\rm  MFT}$, which
is fixed by the condition that the energy density in the stationary
state is the same as in the initial state \fr{Eq:InitialEnergy}, i.e. 
\be
e_0=\lim_{L\to\infty}\frac{\langle H_{\rm MFT}\rangle_{\rm MFT}}{L}\ .
\label{Edens}
\ee

The various self-consistency equations are most easily solved in
momentum space. As stated above it is sufficient to work in
the Neveu-Schwarz sector for even system sizes $L$, so that
\be
c_k \equiv \frac{1}{\sqrt{L}} \sum_m e^{ikm} c_m \ , \qquad k \in \left\{2\pi
\frac{n+1/2}{L} \ , \ n=-\frac{L}{2},\dots, \frac{L}{2}-1\right\}  \ .
\ee
The mean-field Hamiltonian then becomes
\begin{align}
 H_{\rm MFT} &= \sum_{k>0}A_k (c_k^\dag c_k-c^\dag_{-k}c_{-k})+iB_k (c_k^\dag c_{-k}^\dag)-iB^*_k(c_{-k}c_k) + {\rm const}\ ,  \label{Eq:H_MFT_k}\nn
 A_k &= 2\sum_{a=0}^2 J_{\rm Eff}^{(a)}\cos ak \ , \qquad B_k = 2\sum_{a=1}^2 \Delta_{\rm Eff}^{(a)}\sin ak \ . 
\end{align}
We remark that in equilibrium not just the $t_a$ but also the $\Delta_b$ are in fact real despite the absence of a unitary symmetry enforcing this, see Appendix \ref{app:reality}. This in turn makes it
possible to diagonalize the Hamiltonian by a one-parameter Bogoliubov transformation
\begin{align}
    b_\kappa(k)=& \cos\frac{\theta_\kappa(k)}{2}c(k) - i \sin\frac{\theta_\kappa(k)}{2}c^\dag(-k) \ , \qquad
     e^{i\theta_\kappa(k)} = \frac{A_k-iB_k}{\sqrt{A_k^2+B_k^2}} \ ,
    \label{Eq:Bogoliubov}
\end{align}
which gives \footnote{Here we write $|B_k|^2$ which gives the correct dispersion for complex $B_k$, as it will be  out-of-equilibrium, although the form of the required canonical transformation in \fr{Eq:Bogoliubov} will be more complicated. }
\begin{align}
H_{\rm MFT}&=\sum_{k>0}
\varepsilon_\kappa(k)b^\dagger_\kappa(k)b_\kappa(k)+{\rm const}\ ,\qquad
\varepsilon_\kappa(k) = \sqrt{A_k^2+|B_k|^2}.
\label{Eq:MFT_Dispersion}
\end{align}

The self-consistency conditions on the mean-fields are given by
calculating the expectation values using \fr{Eq:SelfConsistent}
\begin{align}
t_a =& \frac{1}{L}\sum_k e^{-iak}\langle c_k^\dag c_k \rangle_{\rm
  MFT}
= \frac{1}{L}\sum_{k>0} \cos ak \left(1-\cos \theta_\kappa(k)\tanh\frac{\beta_{\rm MFT}\varepsilon_\kappa(k)}{2}\right) \ ,     \label{Eq:SelfConsistentt}\\
\Delta_a =& \frac{1}{L}\sum_k e^{-iak}\langle c_k^\dag c_{-k}^\dag\rangle_{\rm MFT} = \frac{1}{L}\sum_{k>0} \sin ak \sin \theta_\kappa(k) \tanh \frac{\beta_{\rm MFT}\varepsilon_\kappa(k)}{2} \ , 
     \label{Eq:SelfConsistentDelta}
\end{align}
while the equation fixing the effective temperature
\fr{Eq:InitialEnergy} takes the form 
\begin{align}
& 4\kappa \left( (t_1+ \Delta_1)^2-(t_0-1/2)(t_2+\Delta_2) \right)_{\kappa=0}
  +h-\frac{1}{L}\sum_{k>0} \varepsilon_{\kappa=0}(k)\tanh\frac{\beta_i
    \varepsilon_{\kappa=0}(k)}{2}\nn
  & = E_0+J^{(0)}_{\rm Eff} - \frac{1}{L}\sum_{k>0} \varepsilon_\kappa(k)\tanh\frac{\beta_{\rm MFT} \varepsilon(k)}{2} .
    \label{Eq:SelfConsistentBeta}
\end{align}

The initial energy density given by the left hand side of
\fr{Eq:SelfConsistentBeta} is a constant for fixed values of $\kappa,
h$, however the right-hand side depends upon the values of the mean-fields and thus this equation must be solved self-consistently
along with the other conditions on the mean-fields.

Eqs \fr{Eq:SelfConsistentt}-\fr{Eq:SelfConsistentBeta} need to be
solved numerically, where the Bogoliubov angles are defined by Eq
\fr{Eq:Bogoliubov} and Eq \fr{Eq:Parameters}. The solutions
can be directly compared to numerical results obtained in
Ref.~\cite{Haldar_2021} via a numerical linked cluster expansion \cite{Rigol2014Quantum,Rigol2016Fundamental}.
\begin{figure}[ht]
\centering
     \subfloat{\includegraphics[scale=1.0]{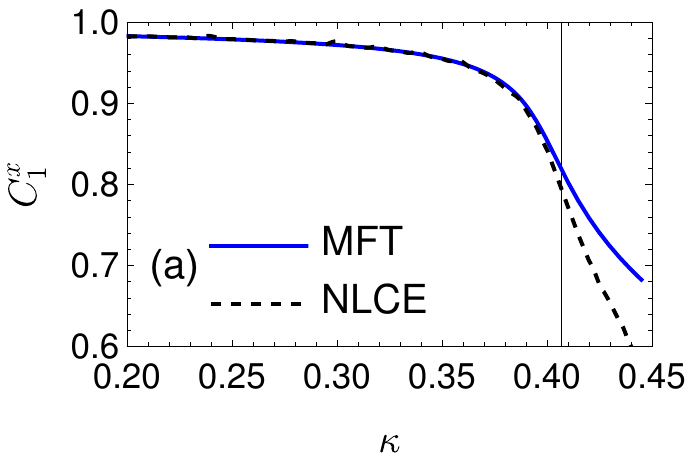}}  \quad 
     \subfloat{\includegraphics[scale=1.0]{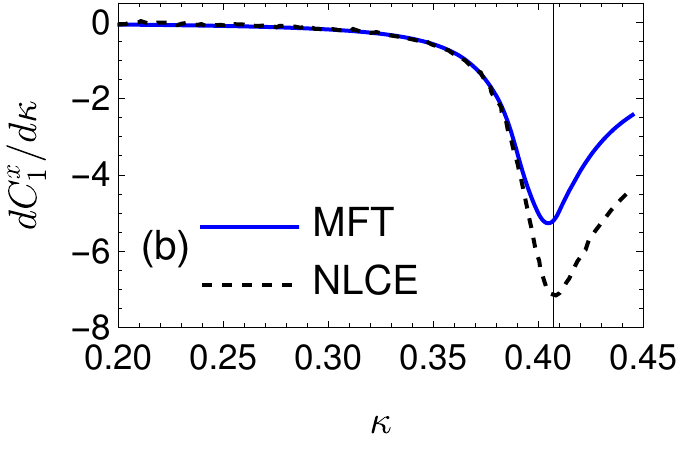}}
\caption{(a) $C^x_1=2(t_1+\Delta_2)$ in the thermal state reached
at late times after a quench from the TFIM ground state at $h=0.2$ as a function of $\kappa$. 
The solid blue line is the result obtained from our self-consistent
mean-field theory and the dashed black line shows numerical linked
cluster expansion (NLCE) results extracted from \cite{Haldar_2021}.
(b) Same comparison as (a) but for $\chi_1=\partial_\kappa
C^x_1(\kappa)$. The vertical lines indicate $\kappa_c$.}
\label{fig:C1Plots}
\end{figure}
\begin{figure}[h]
	\centering
	\includegraphics[scale=1.0
	]{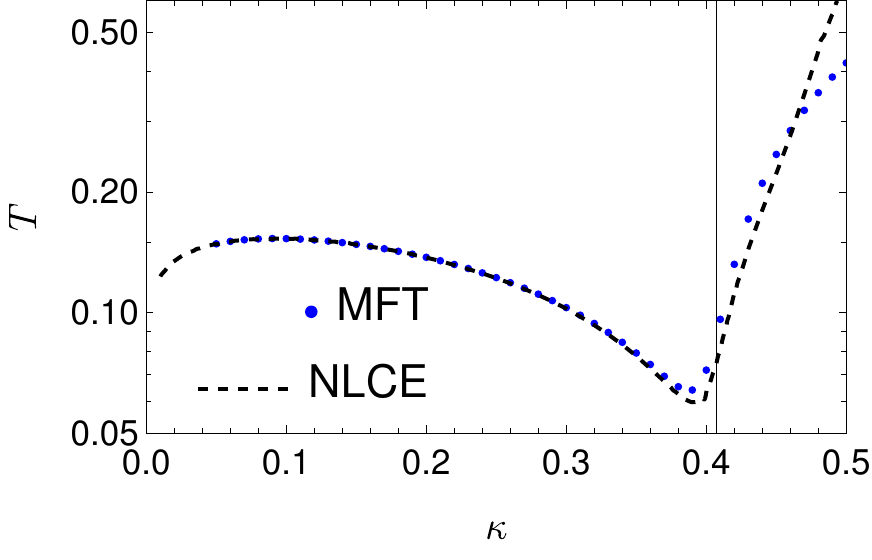} \ 
	\caption{Comparison of $T=\beta^{-1}_{\rm MFT}$ to effective
		temperatures reported in \cite{Haldar_2021}. The dashed black curve shows the
		NLCE results reported in Fig.~8 of \cite{Haldar_2021}, while the blue
		data points are the values found by our self-consistent mean-field
		theory. The vertical line indicates $\kappa_c$. }  
	\label{fig:betaPlot}
\end{figure}
In Fig.~\ref{fig:C1Plots} we plot the
mean-field results for the longitudinal nearest-neighbour correlator
\be
C^x_1 \equiv \langle \sigma^x_i \sigma^x_{i+1} \rangle  = 2(t_1+\Re \Delta_1) \ , 
\label{C1x}
\ee
in the (thermal)
steady state following a quench from the ground state of the TFIM with $h=0.2$ along with the susceptibility ${\rm d}C^x_1/{\rm
  d}\kappa$ defined using an ensemble of quenches. We see that the
agreement of our mean-field analysis with the numerical results of
Ref.~\cite{Haldar_2021} is excellent up to fairly large values of
$\kappa$. We observe similarly good agreement with the transverse magnetization $m^z\equiv \langle \sigma^z_j \rangle$ and the next-nearest neighbour longitudinal correlator $ C^x_2 \equiv \langle \sigma^x_i \sigma^x_{i+2} \rangle$. 
In Fig.~\ref{fig:betaPlot} we compare the self-consistent inverse
temperature $\beta_{\rm MFT}$ to numerical results of
Ref.~\cite{Haldar_2021}. We observe excellent agreement essentially
over the full range of $\kappa$ considered.

Given the good agreement with state-of-the-art numerical results
we conclude that our self-consistent fermionic mean-field theory provides
a good description of the steady state reached at late times after the
quenches considered.

\subsection{Scaling regime at finite energy densities}
\label{Sec:Cutoff}
\begin{figure}[h]
    \centering
    \includegraphics[scale=0.8]{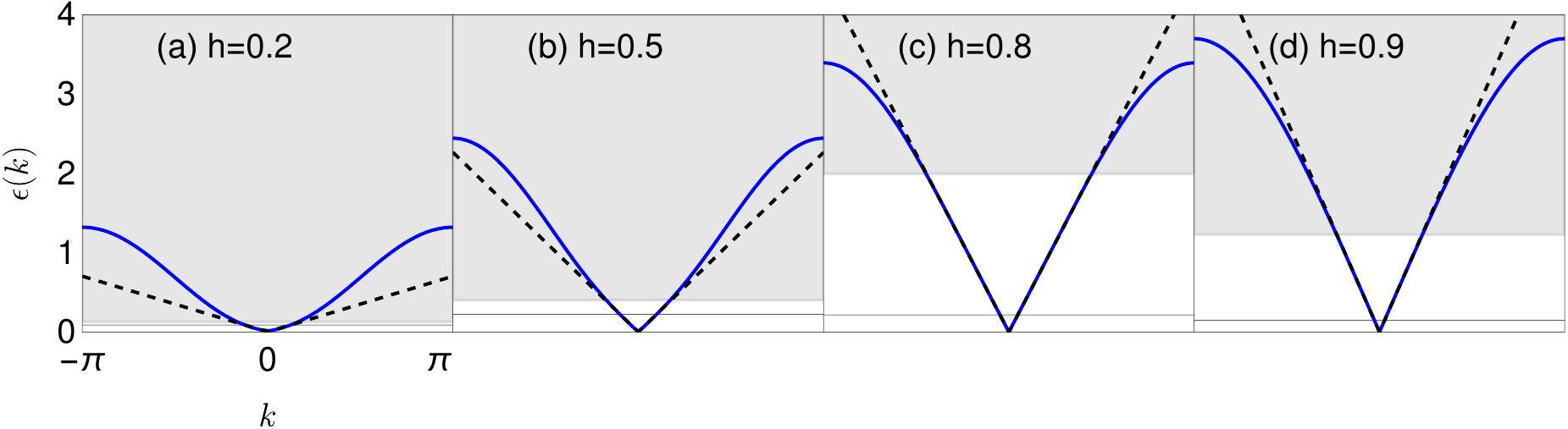}
    \caption{Effective dispersion relations in the steady state following a quench with (a) $h=0.2$, $\kappa=0.407\approx \kappa_c$, (b)
$h=0.5$, $\kappa=0.269\approx \kappa_c$, (c) $h=0.8$, $\kappa=0.114\approx \kappa_c$, (d)
$h=0.9$, $\kappa=0.058\approx \kappa_c$ . The black horizontal line is the effective
temperature $T=\beta_{\rm MFT}^{-1}$. The dashed black line is a fit to
$\varepsilon_{\rm fit}(k)=\sqrt{\varepsilon_\kappa(0)^2+v_{\rm
    fit}^2k^2}$ and the gray shaded region indicates the regime of
energy densities where spectral non-linearities become
significant and corrections to scaling limit behaviour can no longer be expected to be
negligible.}
    \label{fig:scalingcutoff}
\end{figure}

The key objective of Ref.~\cite{Haldar_2021} was to establish that
quantum quenches can be used to locate the positions of quantum phase
transitions in some parameter space. An important question is to
what extent the observed signatures are indeed associated with the
scaling behaviour induced by the proximate quantum critical point. To
answer this question by purely numerical methods would require the
analysis of the long-distance behaviour of correlation functions or
entanglement entropies of large sub-systems, in order to ascertain
whether they display scaling behaviour characteristic of the proximate
quantum critical point. Our mean-field theory gives us a much simpler
way of answering this question: as the field theory describing the
quantum critical point is a gapless relativistic Majorana fermion the
scaling regime extends at most to energies per particle at which the mean-field
dispersion is still to a good approximation linear. These considerations set an energy cut-off for the field theory.
In Fig.~\ref{fig:scalingcutoff} we plot
the mean-field dispersion relation \fr{Eq:MFT_Dispersion} and compare
it to the respective effective temperatures.
We see from Fig.~\ref{fig:scalingcutoff}(c,d) that when $h$ is close to $1$ and $\kappa$ small, the scale over which the dispersion is linear is much
larger than the effective temperature. This implies that for these
quenches the steady state is in fact in the scaling regime of the
Ising transition and properties of the underlying quantum critical
point are readily accessible.

By contrast in Fig.~\ref{fig:scalingcutoff}(a,b) we show the mean-field dispersion relation \fr{Eq:MFT_Dispersion} in the steady state for quenches with
small $h$ and large $\kappa$ can be fitted with a relativistic dispersion only for a small energy window.
Here the scale over which the Majorana dispersion is linear is very
small and of the same order of magnitude as the effective
temperature. This means that for these quenches the steady state is
\emph{outside} the scaling regime of the Ising transition, and so we can't actually glean any useful
information about the underlying quantum critical point using
quench dynamics.

We expect the fact that the cut-off decreases for smaller values of $h$ to be an accurate prediction of the mean-field theory presented here in light of the good agreement with the numerics seen in Fig.~\ref{fig:C1Plots}. The point that the energy density needs to be sufficiently below the
cut-off scale of the quantum critical point one is trying to probe is
of course both obvious and very general.

\section{Self-consistent time-dependent mean-field theory (SCTDMFT)}
\label{Sec:TDMFT}
Following
Refs~\cite{boyanovsky1998evolution,sotiriadis2010quantum,vannieuwkerk2019self,lerose2019impact,Collura20Order,vannieuwkerk2020on,vannieuwkerk2021josephson}
we now turn to the dynamics after our quantum quenches in the
framework of a self-consistent time-dependent Gaussian
approximation. This amounts to considering time evolution with a
time-dependent mean-field Hamiltonian
\begin{align}
H_{\rm MFT}(t)= \sum_i \sum_{a=0}^2 \left\{J_{\rm Eff}^{(a)}(t)(c^\dag_i c_{i+a} +{\rm hc})+ (\Delta_{\rm Eff}^{(a)}(t) c^\dag_i c^\dag_{i+a}+{\rm hc}) \right\} +E_0(t) \ , \label{Eq:MFT_Ht}
\end{align}
where the time-dependent couplings are given by the time-dependent
analogs of \fr{Eq:Parameters}, i.e.
\begin{align}
  t_a(t)&=\text{Tr}\Big(\rho_{\rm MFT}(t)c_j^\dagger
  c_{j+a}\Big)\ ,\quad a=0,1,2\ ,\nn
  \Delta_b(t)&=\text{Tr}\Big(\rho_{\rm MFT}(t)c^\dagger_j c^\dagger_{j+b}\Big)\ ,\quad
  b=1,2\ ,\nn
  \rho_{\rm MFT}(t)&= \left\{  \mathcal{T} e^{-i\int_0^t   H_{\rm MFT}(t') {\rm d}t'}\right\} \rho(t=0) \left\{ \mathcal{T} e^{-i\int_0^t   H_{\rm MFT}(t') {\rm d}t'} \right\}^\dag \ .
\end{align}
Here $\mathcal{T}$ denotes time ordering; the initial density matrix $\rho(t=0)$ \fr{rhoinitial} is by
construction Gaussian and concomitantly so is $\rho_{\rm
  MFT}(t)$. This is the essence of the SCTDMFT, which by construction is expected to work best at short times. This is because it is
based on the assumption that all higher cumulants vanish, which is
strictly true at time $t=0$. At short times the higher cumulants will
become non-zero, but their growth is expected to be slow for small
$\kappa$.

At late times SCTDMFT is not expected to work well in general
\cite{Bertini15Prethermalization,Bertini16Thermalization} and in some
models is known to describe relaxation towards a ``prethermalization
plateau''\cite{moeckel2009real,moeckel2010crossover,essler2014quench}. This
can be described by a deformation of a Generalized Gibbs ensemble
rather than the Gibbs ensemble that describes the thermalized steady state.
The studies of Refs
\cite{Bertini15Prethermalization,Bertini16Thermalization} in which the
accuracy of the SCTDMFT was assessed in detail focused on initial
states that led to non-trivial dynamics even in the absence of
interactions. In such settings the prethermalization plateau is
generally well-separated from the corresponding thermal expectation
values, and as a result SCTDMFT cannot describe the late-time
behaviour even qualitatively. 

However, following Ref.~\cite{Haldar_2021} we here consider initial
states for which there is no dynamics without quenching the
interaction strength. This is the scenario investigated by \cite{moeckel2009real,Nessi2014Equations,Nessi2015glass}. As a result the system still prethermalizes, but
now for local quantities the prethermalization plateau is quite close to the
corresponding thermal expectation values. Given that the prethermalization plateau is reached at early times where SCTDMFT works well, the expectation is that the difference to the Gibbs value at small $\kappa$ is of order ${\cal O}(\kappa^2)$. Hence, SCTDMFT can be viewed
as providing a reasonable account of the late time behaviour (as long
as the quench parameter $\kappa$ is small enough, which must be checked numerically). We elaborate on this point
below by comparing the late-time behaviour in SCTDMFT to the expected
equilibrium values.

As a consequence of the translation invariance of the problem 
the time-evolved Gaussian density matrix $\rho_{\rm MFT}(t)$ is fully
characterised by the two momentum space two-point averages 
\be
\Tilde{t}_k(t)={\rm Tr}\Big(\rho_{\rm MFT}(t)\ c_k^\dag c_k \Big)\ ,\qquad
\Tilde{\Delta}_k(t) ={\rm Tr}\Big(\rho_{\rm MFT}(t)\ c_k^\dag c_{-k}^\dagger \Big).
\ee
The self-consistent equations of motion
for these $k$ space two-point functions can be obtained using the
Heisenberg equations of motion associated to the (now time-dependent) analog of the momentum space Hamiltonian \fr{Eq:H_MFT_k}. The result is 
\begin{align}
\odiff{\Tilde{\Delta}_k(t)}{t} =& 2iA_k(t)\Tilde{\Delta}_k(t)+B_k^*\Big[1-2\Tilde{t}_k(t)\Big] \nn 
\odiff{\Tilde{t}_k(t)}{t} =& 2\Re\Big(B_k(t) \Tilde{\Delta}_k(t)\Big)   \ ,
\label{Eq:1DEoMs}
\end{align}
where
\begin{equation}
    A_k = 2\sum_{a=0}^2 J_{\rm Eff}^{(a)}(t)\cos ak \ , \qquad B_k = 2\sum_{b=1}^2 \Delta_{\rm Eff}^{(b)}(t)\sin ak \ .
\end{equation}
We now integrate the equations \fr{Eq:1DEoMs} using a second-order midpoint scheme
with a timestep of $10^{-3}$, which we choose to ensure that the
mean-fields are converged with respect to the timestep. At each timestep we must update the real space mean-fields $t_a$ and $\Delta_b$ using $\Tilde{t}_k$ and
$\Tilde{\Delta}_k$
\be
t_a=\frac{1}{L}\sum_k\Tilde{t}_k(t)e^{-ika}\ ,\qquad
\Delta_b=\frac{1}{L}\sum_k\Tilde{\Delta}_k(t)e^{-ikb}\ .
\ee
Physical
quantities such as spin-spin correlation functions can then be
calculated in terms of (sums of products of) the fermionic two-point functions.  
%
%
\begin{figure}[h]
\centering
\includegraphics[scale=1.0]{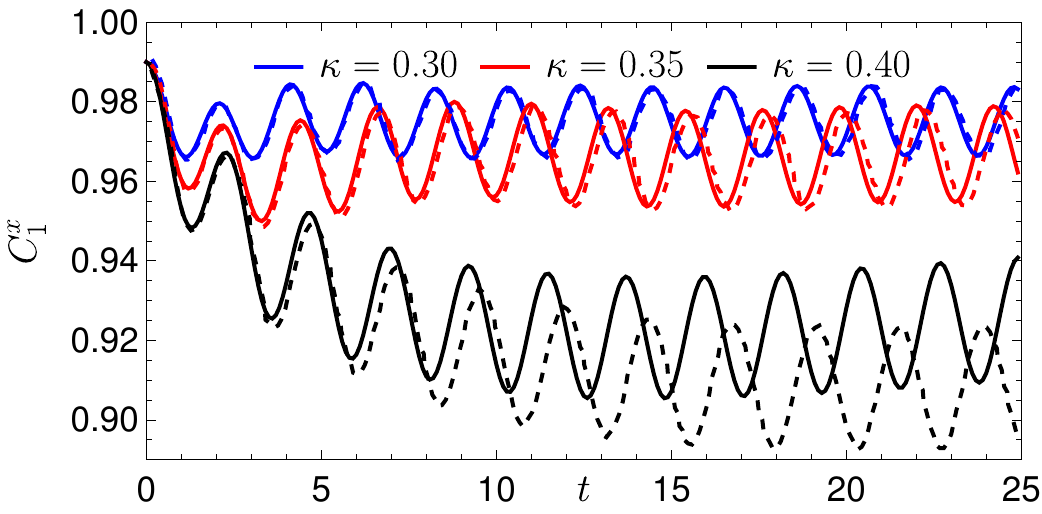}
\caption{Comparison of SCTDMFT results for $C^x_1(t)$ to iTEBD results
taken from \cite{Haldar_2021} for a quench from the ground state at
$h=0.2, \kappa=0$ to $\kappa>0$. Here the solid lines are SCTDMFT 
results for $L=2000$ and the dashed lines in the respective color
are iTEBD. The agreement is seen to be very good except for near the
critical point ($\kappa_c \approx 0.407$).} 
\label{fig:SCTDMFTvsiTEBD}
\end{figure}
\subsection{Short and intermediate-time behaviour of local correlation functions}
In Fig.~\ref{fig:SCTDMFTvsiTEBD} we compare the results of the above
SCTDMFT approximation to iTEBD results taken from
\cite{Haldar_2021}, which are believed to be essentially numerically
exact. For small values of $\kappa$ compared to the critical value $\kappa_c$ we find excellent
agreement over the entire time range accessible to iTEBD. For larger
values of $\kappa$ the agreement is still very good at short times,
but gets worse at late times. 

\begin{figure}[ht]
\includegraphics[scale=1.0]{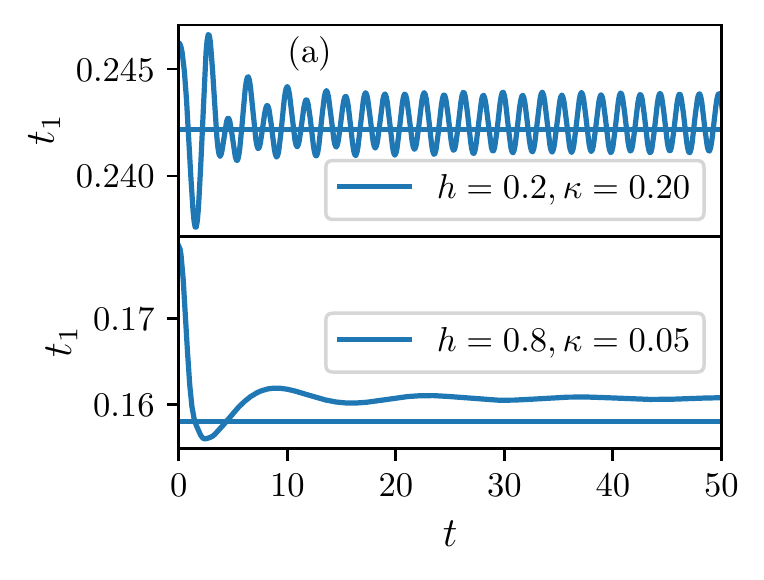}\quad
\includegraphics[scale=1.0]{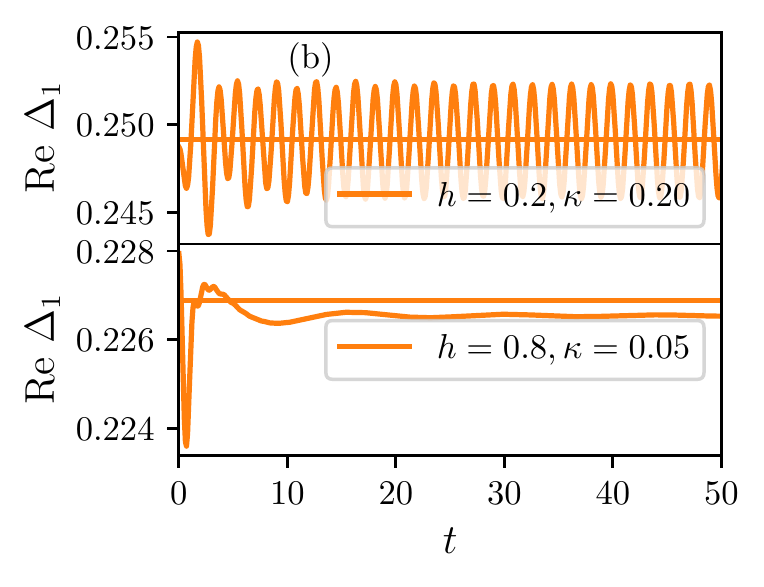}
\caption{Nearest neighbour fermion two-point functions $t_1(t)$, $\Re
\Delta_1(t)$ after quenches from the ground state of $H(h,\kappa=0)$ with
$h=0.2$ and $h=0.8$ to $H(h,\kappa)$. Horizontal lines indicate the stationary
values found in Section \ref{Sec:SCMFT}. }
    \label{fig:MFs_realtime}
\end{figure}

While Ref.~\cite{Haldar_2021} focused on spin correlations, the time
evolution of the fermionic two-point functions is of interest as well,
in particular in relation to the question of detecting topological
transitions by quench dynamics. In Fig.~\ref{fig:MFs_realtime} we
present results obtained by SCTDMFT for $t_1(t)$ and $\Re(\Delta_1(t))$
following quenches from the ground state of $H(h,\kappa=0)$ with
$h=0.2,0.8$ to $\kappa=0.05,0.20$.
We observe the following:
\begin{itemize}
\item{}For quenches with small transverse fields $h$ there are
persistent oscillations around a constant value, which is in good
agreement with the corresponding expectation value after
thermalization. 
\item{}For quenches at large fields $h$ there are no long-lived
oscillations. Instead the expectation values relax to stationary
values that differ from the ones predicted (in mean-field theory) by thermalization, by an
amount that we find scales as ${\cal O}(\kappa^2)$ as expected. 
\end{itemize}
An explanation of the oscillatory behaviour is provided below
in section~\ref{Subsec:Oscillations}.

\begin{figure}[ht]
    \centering
    \includegraphics[scale=1.0]{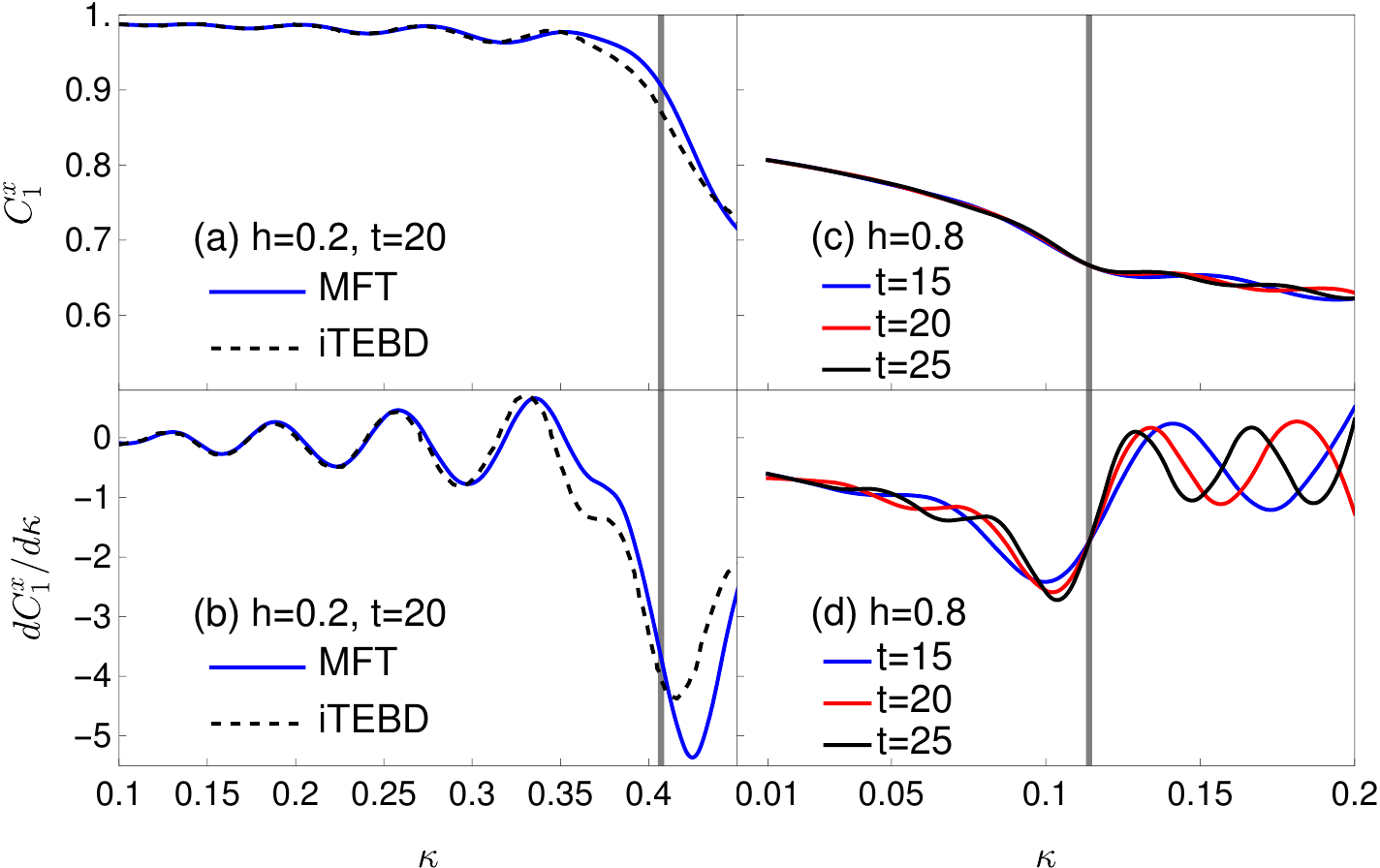}
    \caption{Performing quenches from $H(h,0)$ to $H(h,\kappa)$ we build a picture of observables as a function of final $\kappa$. (a-b) Comparison with iTEBD data taken from \cite{Haldar_2021} for $h=0.2$ ($\kappa_c\approx 0.407$, indicated by thick gray line). (c-d) Equivalent calculation at $h=0.8$ ($\kappa_c \approx 0.114$). All quenches done starting from the ground state for system size $L=2000$.}
    \label{fig:timeslice}
\end{figure}

As suggested in \cite{Haldar_2021}, a signature of the proximate
quantum phase transition can be obtained by processing data
for the expectation value of a local observable for an ensemble of
quenches at a fixed time $t$ after the quench. In
Fig.~\ref{fig:timeslice} we show results for $C_1^x(t)$
and $dC_1^x(t)/d\kappa$ for an ensemble of quenches starting in the
ground state of $H(h,\kappa=0)$ and quenching to $H(h,\kappa)$ for $h=0.2,0.8$ and a wide range of $\kappa$ values. 

In Fig.~\ref{fig:timeslice}(a-b) we find very good agreement between our SCTDMFT results and the iTEBD
simulations of Ref.~\cite{Haldar_2021} for $h=0.2$ and in Fig.~\ref{fig:timeslice}(c-d) we show the results for $h=0.8$. The generalized susceptibility
${\rm d}C_1^x/{\rm d}\kappa$  in Fig.~\ref{fig:timeslice}(b,d) shows a strong dip even at the relatively early time $t=20$ around the critical
value $\kappa_c$. Intuitively one expects that the reason for this
strong response to the varying post-quench parameters is that the
correlation length at time $t=20$ is already large and the system
``feels'' the proximity of the QPT; this implies a large correlation length and consequently a strong linear response of the system, reflected in the dips in generalized susceptibilities. We return to this point in the
next section where, in Fig.~\ref{fig:SxSxCorrs}, we extract correlation lengths for the non-equilibrium state of the system following the quench for $h=0.8$ and find that the correlation length has grown from $\xi\approx 1.9$ at $t=0$ to $\xi\approx 12 $ at $t=20$. 
Conversely, in cases where the correlation length is
short we do not expect the susceptibility to be large. This is indeed
the case for small values of $\kappa$ in Fig.~\ref{fig:timeslice}.
\begin{figure}[ht]
\includegraphics[scale=1.0]{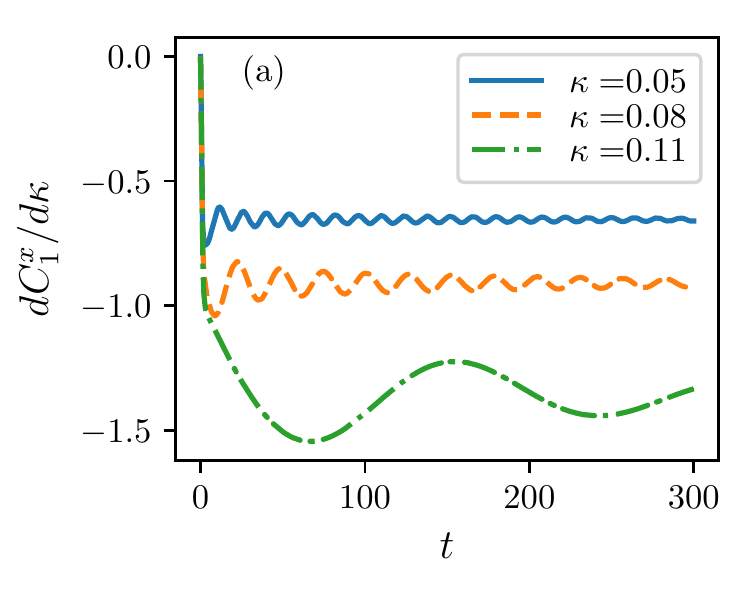}\quad
\includegraphics[scale=1.0]{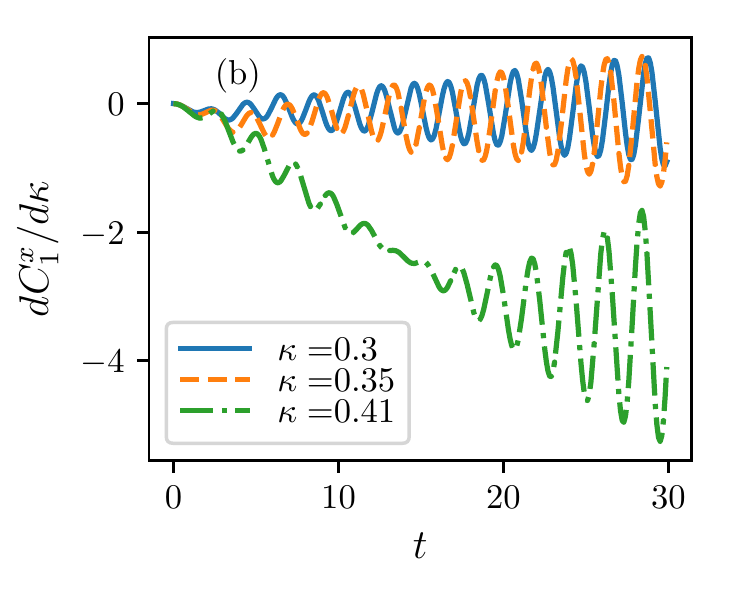}
\caption{Short time dynamics of the generalized susceptibility for quenches
  from an initial thermal state with $\beta=2.0$ and (a) $h=0.8$ ($\kappa_c\approx 0.114$)
  and (b) $h=0.2$ ($\kappa_c\approx 0.407$) on a system with $L=2000$.}
    \label{fig:Short_times}
\end{figure}
 In Fig.~\ref{fig:Short_times} we show the time evolution of the
generalized susceptibilities. 
Fig.~\ref{fig:Short_times} shows , for two values of $h$, quench data for various $\kappa$, including near the critical value $\kappa_c$. For $\kappa$ far from $\kappa_c$ we observe a quick relaxation to a plateau, whilst for $\kappa$ close to the QPT we observe a longer relaxation time. Fig.~\ref{fig:Short_times}(b) features growing oscillations due a `beat' phenomenon when numerically differentiating between the different quench data with slightly different persistent oscillation frequencies.

\subsection{Growth of the correlation length in time}
As we have noted above, the correlation length grows in time for many of
the quenches we consider. To show this explicitly we focus on the connected order-parameter
two-point function
\be
C^x_{c,\ell}(t)=\underbrace{{\rm Tr}\Big[\rho_{\rm MFT}(t)\ \sigma^x_n
  \sigma^x_{n+\ell}\Big]}_{C_\ell^x(t)}
- \Big({\rm Tr}\Big[\rho_{\rm MFT}(t)\ \sigma^x_n \Big]\Big)^2,
\ee
as it is easier to extract a correlation length for than $\sigma^z_j$. Since the order parameter expectation value is itself difficult to calculate even in the TFIM \cite{Dreyer2021Quantum,granet2022out} we follow Ref.~\cite{bravyi2006lieb} in using the Lieb-Robinson bound
\cite{Lieb1972Finite} to express the connected correlator as
\be
C^x_{c,\ell}(t)=C^x_\ell(t)-C^x_{R}(t)\ ,\quad R\gg v_{\rm max}t,
\ee
where $v_{\rm max}$ is the Lieb-Robinson velocity. In our self-consistent mean-field approximation we can use Wick's
theorem to express $C^x_\ell(t)$ as a block-Toeplitz Pfaffian
\cite{barouch1971statistical}
\begin{align}
C^x_\ell(t) =& \Pf 
    \begin{pmatrix}
        G_0(t) & G_1(t) & \dots& G_{\ell-1}(t) \\ 
        -G_1^T(t) & \ddots & \ddots & \vdots \\ 
        \vdots & \ddots & \ddots & \vdots \\ 
        -G_{\ell-1}^T(t) & \dots & \dots &  G_0(t) 
    \end{pmatrix} \ ,
    \label{Eq:Pfaffians}
\end{align}
where
\begin{align}
    G_n(t) =& 2
    \begin{pmatrix}
        i \Im\Delta_n(t) & \Re(t_{1-n}(t)+\Delta_{1+n}(t))-\frac{1}{2}\delta_{0,n+1} \\ 
        - \Re(t_{1-n}(t)+\Delta_{1-n}(t))+\frac{1}{2}\delta_{0,1-n} & i\Im\Delta_n(t) 
    \end{pmatrix} \ . 
\end{align}
We note that if we replace the time-dependent Gaussian density matrix
by a thermal equilibrium state Eq \fr{Eq:Pfaffians} reduces to a
determinant because $\Delta_n \in \R$.

\begin{figure}[ht]
\centering
\includegraphics[scale=0.8]{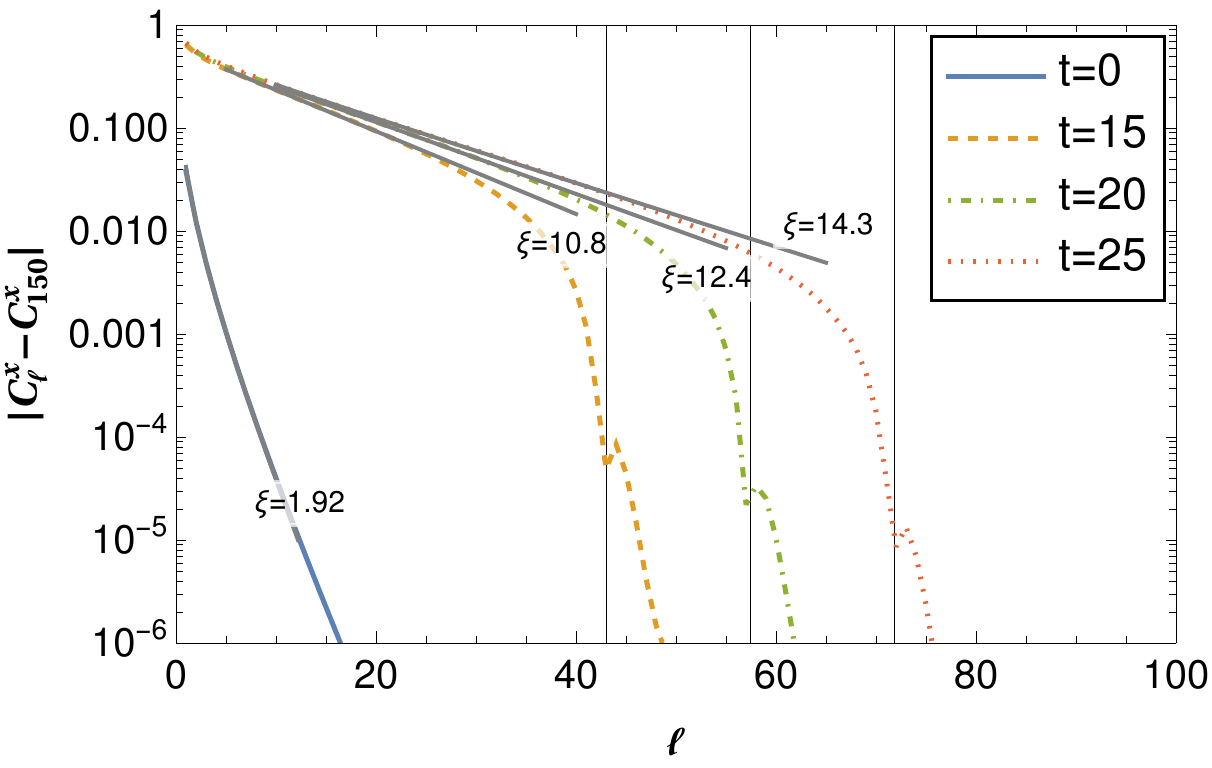}
\caption{Connected order-parameter two-point function $C^x_{c,\ell}(t)$
for a quench from the ground state of the TFIM at $h=0.8$ to the ANNNI with $h=0.8, \kappa=0.11$ ($\kappa_c \approx 0.114$). Vertical lines indicate the lightcone distance at $t=15,20,25$ using the maximal group velocity of the effective dispersion in the steady state. Gray lines indicate fits to functions of the form $C^x_{c,{\rm fit}}=a \ell^{-\nu} \exp(-\ell/\xi)$ where $\xi$ is the fitted correlation length.}
\label{fig:SxSxCorrs}
\end{figure}
In Fig.~\ref{fig:SxSxCorrs} we show the connected order-parameter
two-point function for a quench from the ground state of the TFIM with $h=0.8$ and turning on next nearest neighbour interactions of strength $\kappa=0.11$. In the
initial state the connected correlator displays exponential decay
with a correlation length $\xi(0)\approx 1.9$. Extracting correlation
lengths at $t>0$ is complicated by the fact that the connected
correlator for outside the ``light-cone'' remains unchanged and we are
therefore restricted to separations $\ell < 2v_{\rm max}t$, where
$v_{\rm max}$ is the maximal propagation velocity
\cite{Calabrese2005Evolution,Calabrese2006Time,EsslerQuench2016}. On
the other hand, in order to extract a correlation length $\xi(t)$ we
require that $\ell\gg\xi(t)$. This causes us to be unable to convincingly fit correlation lengths for short times (other than $t=0$ which is an equilibrium state by design), although we obtain relatively good fits to the exponential behaviour at times $t\geq 20$ which show the correlation length has grown to about $\xi(25)\approx 14.3$.

\subsection{Oscillations in the low energy-density regime}
\label{Subsec:Oscillations}
A striking feature seen in Figs.~\ref{fig:SCTDMFTvsiTEBD}, \ref{fig:MFs_realtime},
\ref{fig:Short_times} are the high-frequency oscillations in local
observables for quenches at reasonably small $h$  which do not appear to
decay in time in the mean-field theory. 
These do not occur in quenches in the TFIM and hence seem to be a result of fermion interactions. 
We stress that these oscillations were previously observed in the iTEBD simulations of Ref.~\cite{Haldar_2021} and are not an artifact of the mean-field
approximation. 
Importantly they are observed in quenches that result in
small energy densities compared to the fermion gap, which
puts us in a regime where we are dealing with the non-equilibrium
dynamics of a very dilute gas of fermions. This suggests that
these oscillations could be related to the formation of long-lived
bound states of (pairs of) fermions, \emph{cf.}
Refs~\cite{Lin2017Quasiparticle,KormosRealtime2016,Collura2018Dynamical,Pomponio2019Quasiparticle,Robinson2019Signatures}. 
A simple limiting case in which this bound state formation can be seen is $h=0$. Here excitations are (highly degenerate) domain-wall states, whilst the antiferromagnetic next-nearest neighbour term partially lifts this degeneracy by introducing an energy penalty of $4\kappa$ when the domain-walls are on exactly neighbouring bonds. That is, at $h=0$ the next-nearest neighbour interaction produces a spin-flip (anti-)bound state.
In order to investigate the possibility of these bound states persisting to the non-zero values of $h$ we consider we have determined the spectrum of low-lying excitations of the ANNNI model by exact diagonalization
using the QuSpin\cite{QuSpinPartI17} package on $L=24$ sites. 
These results provide useful information for physical properties at finite
energy densities that are small compared to the excitation gap over
the ground state. 
\begin{figure}[ht]
(a)\includegraphics[scale=0.45]{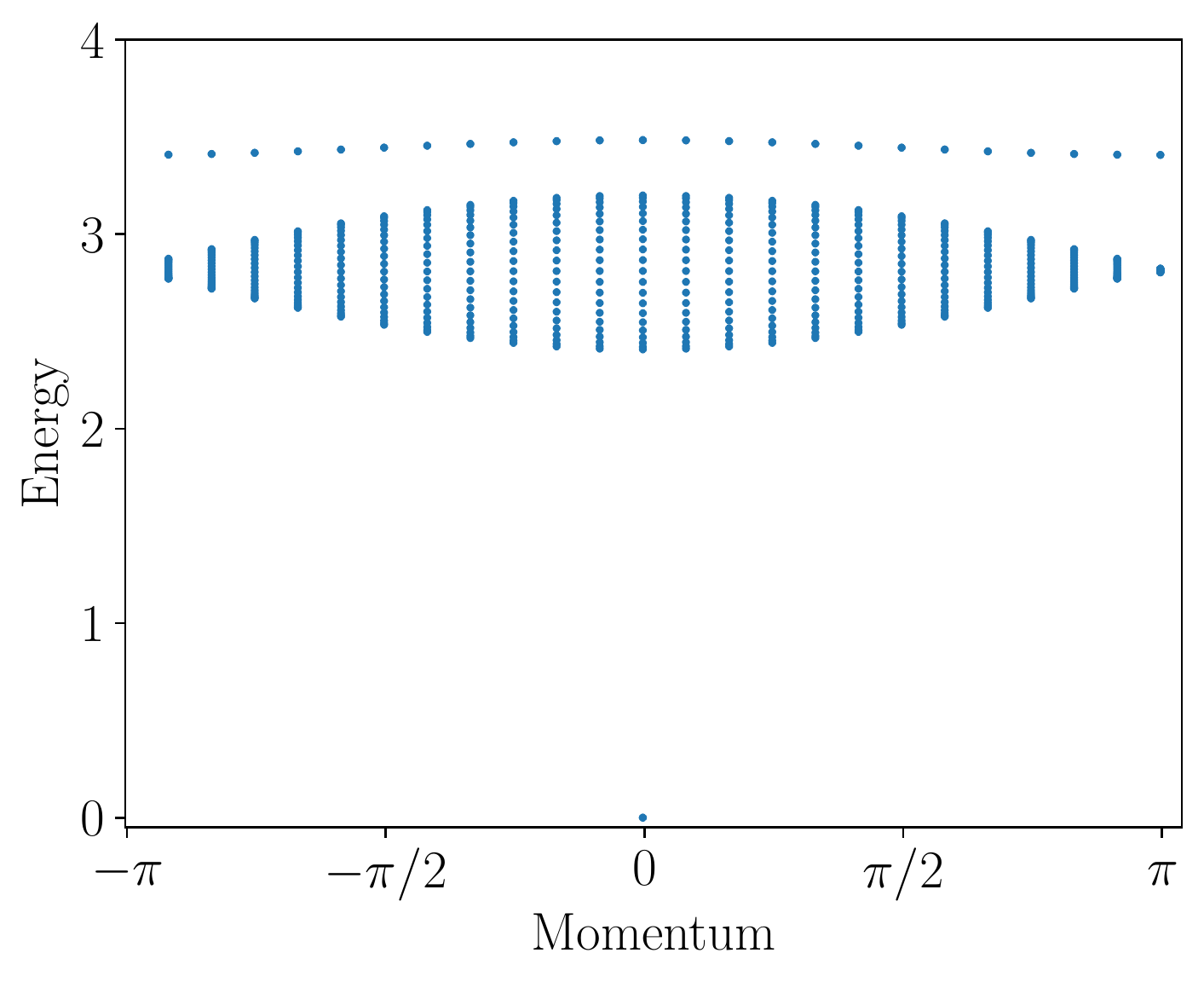}\quad
(b)\includegraphics[scale=0.45]{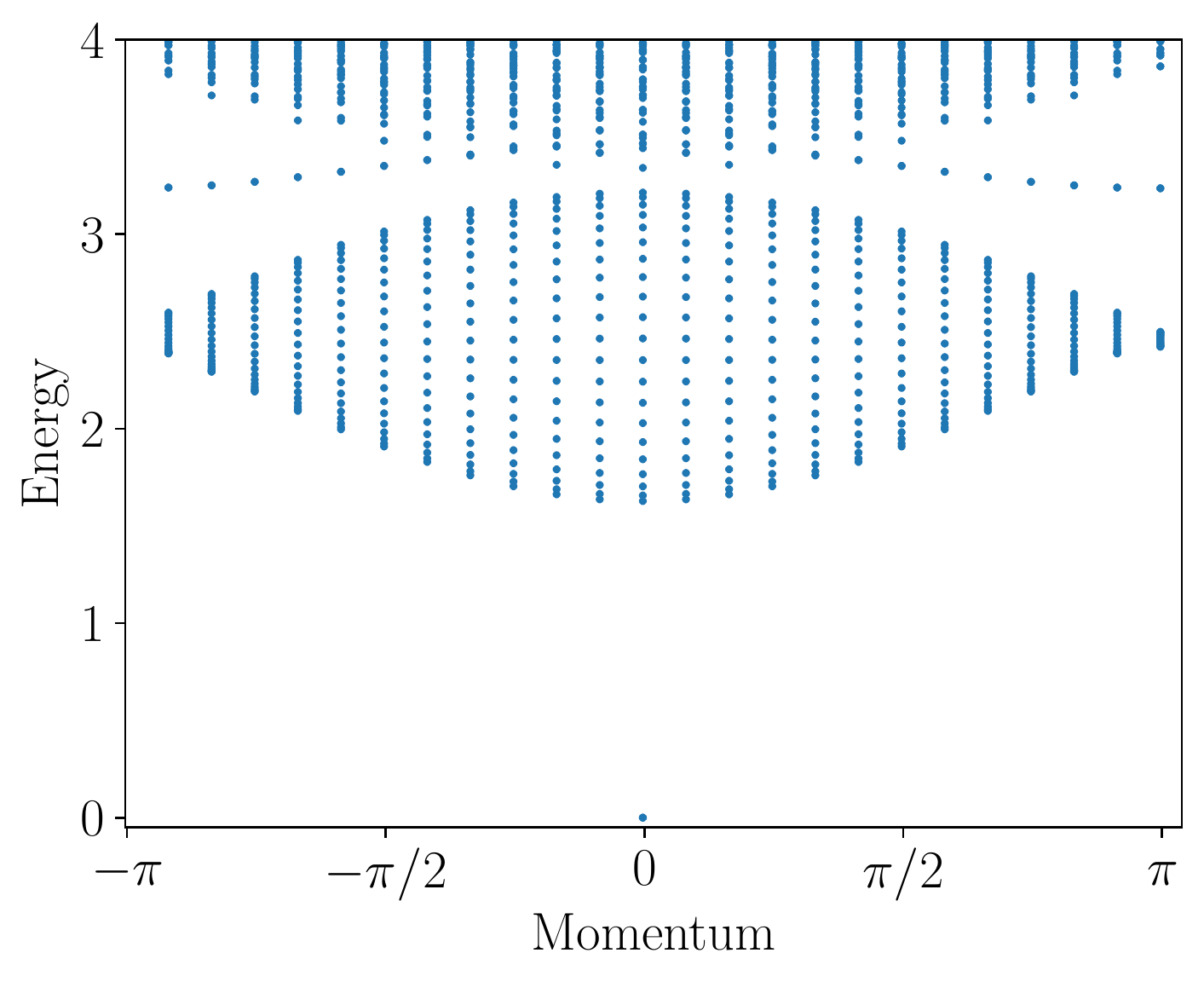}
\caption{Spectrum of the ANNNI Hamiltonian for (a) $h=0.1$,
$\kappa=0.15$ and (b) $h=0.2$, $\kappa=0.2$ from exact
 diagonalisation using QuSpin\cite{QuSpinPartI17} on $L=24$
sites. As physical states have even fermion
parity, the lowest excited states are the two domain-wall continuum and a sharp
bosonic mode corresponding to the anti-bound state. For $h=0.2,
\kappa=0.2$ the four-particle continuum is low enough in energy to
be visible on this scale.}
\label{fig:ANNNI_Spectra}
\end{figure}
As in the ferromagnetic phase of the TFIM the lowest excitations can then be thought of as a continuum of pairs of
ferromagnetic domain-walls. This is indeed observed in the exact
diagonalization results in Fig.~\ref{fig:ANNNI_Spectra}. In addition
we observe a bosonic bound state of two domain-walls that occurs at
energies above the two domain-wall continuum. With regards to the oscillations
observed in local observables after some of our quenches we note the
following:
\begin{itemize}
\item{} The bound state energy at $k=0$ agrees with the oscillation
  frequency observed after the quantum quenches.
\item{} For reasonably large values of $h$ the bound state ceases to exist around $k=0$. It can be seen from a Lehmann representation that only excited states with $k=0$ contribute to the dynamics when performing quenches from translationally invariant states as we do here. As such this is consistent with the fact that when we perform quenches with larger $h$ we do not see persistent oscillations.
\end{itemize}
\label{Subsec:MFP}

An important caveat is that in the quench set-up we are dealing with there is a
small, but finite, energy density above the ground state and thus in the thermodynamic limit the system is in fact at an energy infinitely above what is pictured in Fig.~\ref{fig:ANNNI_Spectra}. There the bound states always
``sit'' on top of multi domain-wall excitations and are not expected
to be stable. However, as the density of domain-walls is very small
the life-time of the bound state can be very large compared to the
time scale we observe in our quenches. We believe that this is indeed
the case.

\begin{figure}
    \centering
    \includegraphics[scale=0.7]{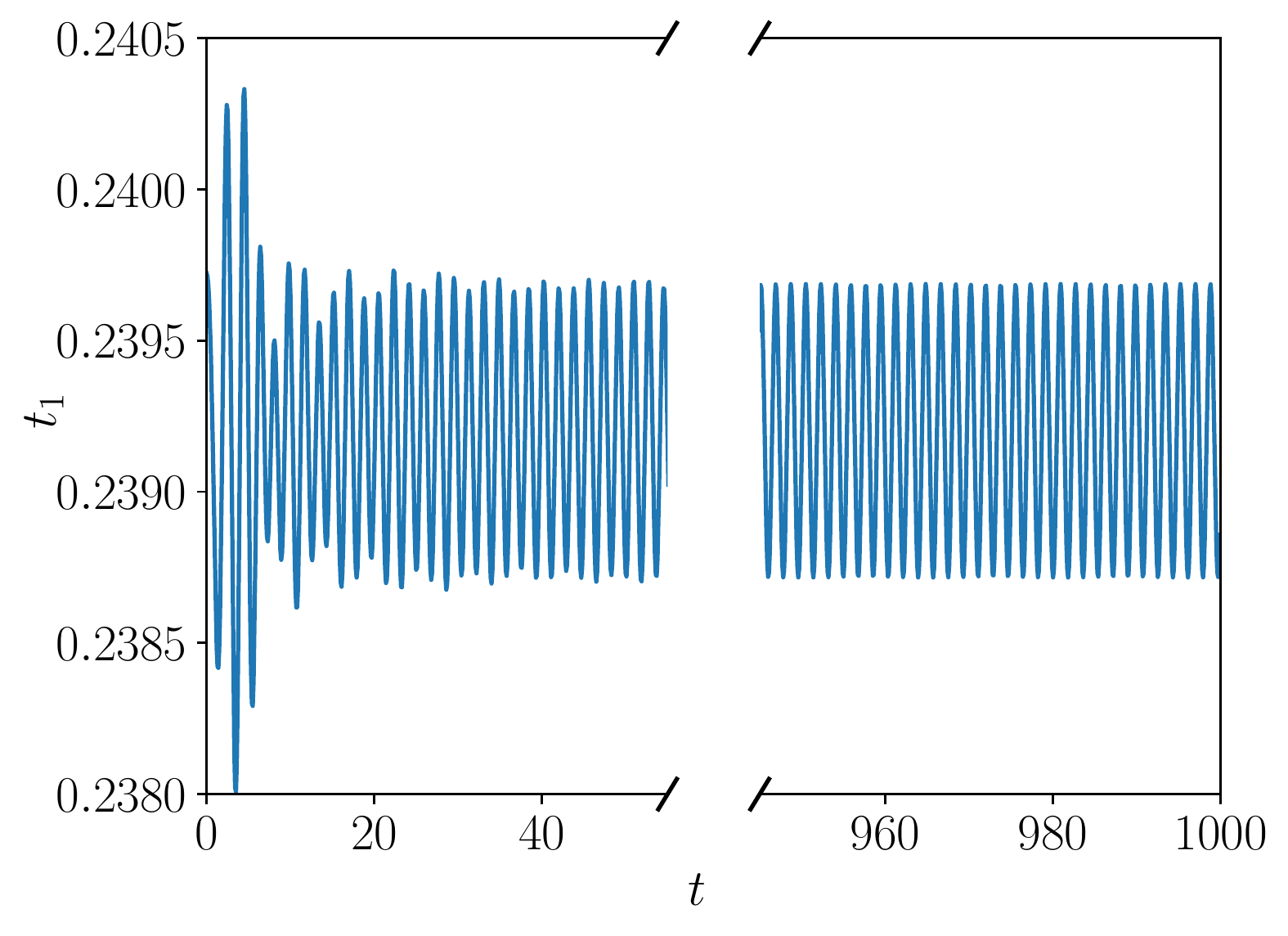}
    \caption{Time evolution of the mean field $t_1$ following a quench from $\beta=2.0$, $h=0.1, \kappa=0.15$.}
    \label{fig:NoDecay}
\end{figure}

A rough estimate of the decay time of the bound states can be obtained by thinking in the quasiparticle picture described above. If there were truly a single bound state then energy and momentum conservation would prevent it from decaying, however the decay is allowed due a background density of domain walls that the bound state may scatter from. A semi-classical approach to compute the scattering time is to reason in terms of the quasiparticle picture \cite{Lin2017Quasiparticle,Rieger2011Semiclassical,Blass16Test}. To do this we introduce the mean-free-path of the domain-walls
\begin{equation}
\lambda_{\rm mfp} = \frac{E_g}{\varepsilon} \ , \label{Eq:MFP}
\end{equation}
where $\varepsilon$ is the energy density \emph{relative to the ground state
after the quench} and $E_g$ the quasiparticle gap. If the mean-free-path is larger than the system size $\lambda_{\rm mfp}>L$, then the state has in expectation fewer than one quasi-particle in the entire system and the system does not require a many-body description and the bound states will have nothing to scatter from. 
Even for thermodynamically large systems however if we consider times less than
\begin{equation}
   2 v_{\rm max} t \lesssim \lambda_{\rm mfp}  \ , 
\end{equation}
where $v_{\rm max}$ is the Lieb-Robinson velocity of the domain-wall excitations, we may consider the bound state quasiparticles as having little interaction with the domain-wall background. 
\begin{figure}[ht]
\centering
\includegraphics[scale=0.7]{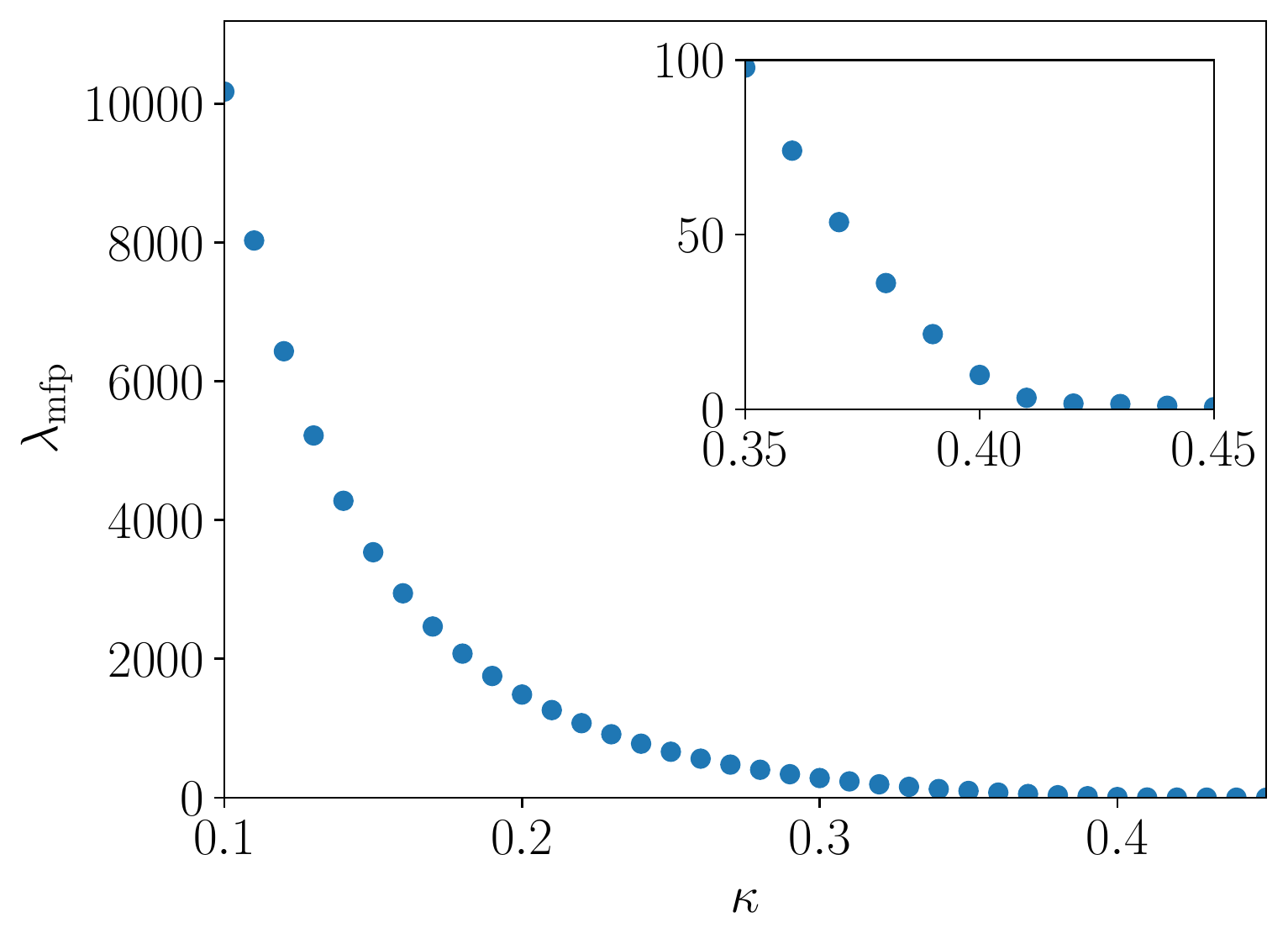}
\caption{Mean free path of the quasiparticles generated by quantum quenches
from the TFIM ground state at transverse field $h=0.2$ to the ANNNI
model with $0.1<\kappa<0.45$ ($\kappa_c\approx=0.407$). } 
    \label{fig:mfp}
\end{figure}
We now estimate all the relevant quantities in the case of interest. The post-quench energy density $e_0$ defined in \fr{Edens} may be calculated using Wick's theorem. The energy density $\varepsilon$ appearing in Eq \fr{Eq:MFP} is however not the $e_0$ of \fr{Edens} but rather one must subtract the ground state energy 
density of the ANNNI, which is not known analytically.
We estimate the latter by exact diagonalization
for $L=18$ sites, for which it is essentially converged. The resulting mean-free-path for quenches from the
ground state of the TFIM with $h=0.2$ to the ANNNI model with
$0.1<\kappa<0.45$ is shown in Fig.~\ref{fig:mfp}. We see that for these
quenches the mean free path is extremely large unless $\kappa$ is very
close to the QPT.
The time range accessible to us in our SCTDMFT analysis is limited by
finite-size effects, which strongly influence observables after the
traversal time $L/(2v_{\rm  max})$
\cite{EsslerQuench2016,Rieger2011Semiclassical, Blass2012Quantum}. To
access very late times without encountering finite-size effects
therefore requires larger system sizes and more memory. In order to
test whether or not the oscillations eventually decay in mean-field
theory we instead change our initial density matrix in a way
that reduces the mean free path, e.g. for a quench with $h=0.1$ and
$\kappa=0.15$ from an initial temperature $\beta=2.0$, we estimate that
the mean free path should be roughly $50$ sites and the scattering
time about $t_s\sim 56$, see Table
\fr{tab:ScatteringTimeCalcs}. Nonetheless there is no visible damping 
in the mean-field theory up to very late times ($t=10^3$), see Fig.~\ref{fig:NoDecay}.  
\begin{table}[ht]
    \centering
    \begin{tabular}{ccccccc}
         $e_0(\beta=2.0)$ & $e_{\rm GS}(h=0.1, \kappa=0.15)$ & $\varepsilon$ & $2E_g$ & $\lambda_{\rm mfp}$ &  $v_{\rm max}$ & $t_s $\\ 
         \hline
       -0.82739 & -0.85295  & 0.02556  & 2.410 & 47.14 & 0.4187 &  56.29
    \end{tabular}
    \caption{Postquench energy density $e_0$ obtained from Eq \fr{Edens}, ground state energy density $e_{\rm GS}$ and two particle gap estimated with ED on $L=20$ sites. Lieb-Robinson velocity is estimated as the maximal group velocity for the dispersion $\epsilon_\kappa(k)$ given in Eq \fr{Eq:MFT_Dispersion} using the values of the mean-fields at $t=100$.}
    \label{tab:ScatteringTimeCalcs}
\end{table}
We conclude that in SCTDMFT the oscillations are \emph{undamped} while we expect in an exact theory they would decay. 

\section{Non-equal time correlation functions}
\label{Sec:NonEqualTimes}
\begin{figure}[ht]
    \centering
    \includegraphics{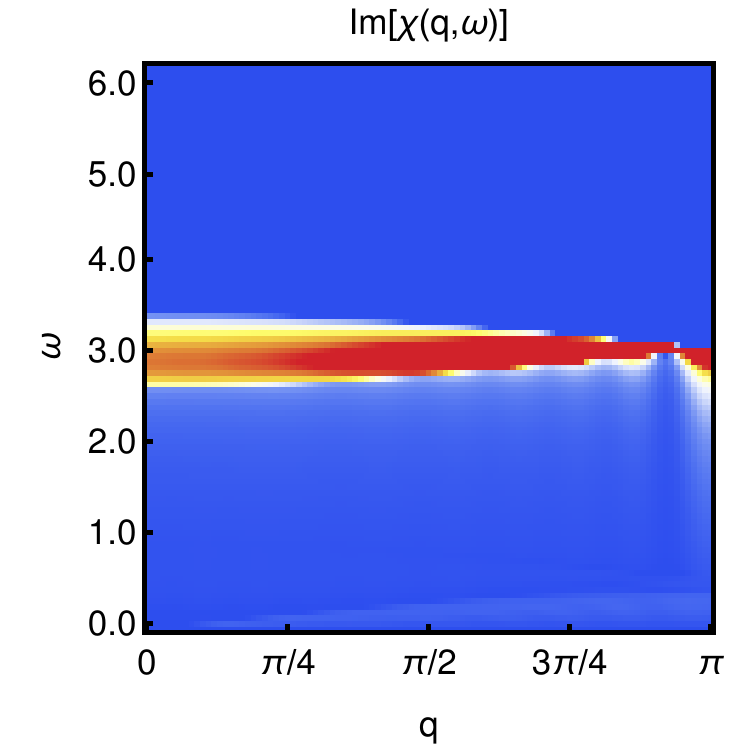}
    \caption{Out-of-equilibrium density-density susceptibility calculated for the mean-field theory with $L=200, h=0.1, \kappa=0.15, \beta=1.0$}
    \label{fig:susceptibility}
\end{figure}

A natural question is whether the existence of a bound state can be
detected more directly in the quench setup. One proposal in the
literature is to use certain Fourier transforms of equal-time
correlation functions \cite{VillaUnraveling2019, VillaLocal2020}, but
these do not provide useful insights in our case. 
In thermal equilibrium it is well established that dynamical response
functions give detailed information about the particle content of the
theory. An obvious question then is to what extent their
non-equilibrium analogs can be used to do the same. In order to
address this question we now determine certain non-equal time
correlation functions in our SCTDMFT. We do not attempt to address the
problem of calculating non-equal time two-point functions of the order
parameter, as this is difficult even for the transverse field Ising
chain itself \cite{essler2012dynamical,granet2022out}. In MFT the
Heisenberg equations of motion for the fermion operators $c_k$ are linear 
\begin{equation}
    \odiff{}{t}c_k(t) = i [H_{\rm MFT}(t),c_k(t)] = -iA_k(t) c_k(t) + B_k c_{-k}^\dag(t) \ , 
\end{equation}
and can be solved by a time-dependent Bogoliubov transformation
\begin{align}
    c_k(t) =& \alpha_k(t) c_k(0) + \beta_k(t) c^\dag_{-k}(0) \ ,  
\end{align}
where the time-dependent coefficients are solutions to 
\begin{align}
    \odiff{\alpha_k(t)}{t} =& -iA_k(t) \alpha_k(t) + B_k(t)\beta_{-k}^*(t) \ , \qquad
    \odiff{\beta_k(t)}{t} = -iA_k(t) \beta_k(t) + B_k(t)\alpha_{-k}^*(t) \ . \label{Eq: Heisenberg EoMs}
\end{align}
As we are dealing with a Gaussian theory all non-equal time
correlation functions are then expressible in terms of the two non-equal
time Green's functions given by  
\begin{align}
    G_k(t,t') = \langle c_k^\dag(t) c_k(t') \rangle = &\alpha^*_k(t) \alpha_k(t') f_k + \alpha_k^*(t) \beta_k(t') g_k \nn &+ \beta_k^*(t) \alpha_k(t') g_k^* + \beta_k^*(t) \beta_k(t')(1-f_{-k}) = G_{-k}(t,t')\ , \\ 
    \Tilde{G}_k(t,t') = \langle c_k^\dag(t) c_{-k}^\dag(t') \rangle = &\alpha_k^*(t) \alpha_{-k}^*(t') g_k + \alpha_k^*(t) \beta_{-k}^*(t')f_{-k} \nn &+ \beta_k^*(t) \alpha_{-k}^*(t') (1-f_k) + \beta_k^*(t) \beta_{-k}^*(t') g^* = -\Tilde{G}_{-k}(t,t') \ , 
\end{align}
where expectation values are always taken with respect to $\rho(t=0)$, i.e. $\langle \mathcal{O}\rangle = \Tr [\rho(t=0) \mathcal{O}]$.
The final equalities hold due to the parity symmetry and
$f_k$, $g_k$ encode the initial conditions
\be
f_k=G_k(0,0) ,\qquad
g_k=\Tilde{G}_k(0,0) \ . 
\ee
As an example of the use of these formulas we consider the
non-equilibrium analog of the density response function
\begin{align}
      \chi_{\rho \rho}(r,t,t') =& \frac{1}{L^2}\sum_{k_1,\dots
        k_4}e^{i(k_1-k_2)r} \langle [c_{k_1}^\dag(t)
        c_{k_2}(t) , c_{k_3}^\dag(t') c_{k_4}(t') ] \rangle\ .
\end{align}
After Fourier transforming in the spatial co-ordinate this takes the
following form in SCTDMFT
\begin{align}
    \Tilde{\chi}(q,t,t') = \frac{1}{L}\sum_k \Big\{&\Tilde{G}_{k}(t,t')\Tilde{G}^*_{k-q}(t',t)-\Tilde{G}_{k}(t',t)\Tilde{G}^*_{k-q}(t,t') \nn
    & + G_{k}(t,t')\left(\alpha^*_{k-q}(t')\alpha_{k-q}(t)+\beta_{k-q}^*(t')\beta_{k-q}(t)\right)  \nn & - \left(\alpha_{k}^*(t)\alpha_{k}(t') + \beta^*_{k}(t)\beta_{k}(t')\right) G_{k-q}(t',t) \Big\} \ . 
\end{align}
We note that $\chi(q,t,t')$ is in principle measurable via
linear-response measurements, see Appendix
\ref{App:LinearResponse}. Employing a Lehmann representation suggests
that spectral properties of the post-quench Hamiltonian should be
inferrable by taking appropriate ``Fourier transforms'' in time. In
practice we consider
\be
\chi_{t_f}(q,\omega)=\int_0^{t_f}dt'\ \Tilde{\chi}(q,t_f,t')\ e^{i\omega t'}.
\ee
The imaginary part of this generalized dynamical susceptibility
is shown in Fig.~\ref{fig:susceptibility} for a quench from $\kappa=0$
to $\kappa=0.15$ and initial inverse temperature $\beta=1.0$. These parameters correspond to those in Fig.~\ref{fig:ANNNI_Spectra}(a), which has a clear bound state, except at elevated temperature. The single time correlation functions still show characteristic oscillations and so we believe there is a bound state still at this increased temperature. In Fig.~\ref{fig:susceptibility} we can clearly
identify the continuum of two domain-wall excitations but there is no
evidence for a bound state above it. In order to capture the latter
one has to go beyond the SCTDMFT.

\section{Conclusion}
We have formulated both equilibrium (at finite energy density) and time-dependent mean-field descriptions for quantum
quenches in the ANNNI model starting from a Gaussian state. 
We first used this to compute properties of the expected stationary state following a quantum quench, assuming that the system looks thermal again at late times and then used the time-dependent formulation to probe the approach to stationarity.
Comparisons in both the stationary and time-dependent cases with the numerical results of Ref.~\cite{Haldar_2021} show
that this simple description is
 fairly accurate even for
large next-nearest neighbour interactions close to the critical
value. Importantly it fully reproduces the signatures of the
equilibrium phase transition previously found numerically. Our 
approach makes it clear that the observed signatures are associated with
the growth of the correlation length following a quantum quench and 
sheds light on the applicability of this mechanism for detecting
quantum phase transitions in general. 
Our analysis makes it clear that to truly see quantum critical behaviour the quench must not produce an energy density higher than the cutoff for the critical scaling theory, however the appearance of soft modes can still produce signatures of the phase transition in the susceptibilities for one point functions.
Our theory is based on a
fermionic description with a topological transition and so it is clear
that topological as well as conventional transitions may be detected
in this manner. Moreover, we give an explanation for a potentially
puzzling feature of the real time dynamics, namely long-lived
oscillations, by showing that the oscillation frequency is the
mass of a bound state in the interacting theory. 

Finally, we showed that self-consistent time-dependent mean-field theory can partially capture the presence of this bound state due to the characteristic oscillations caused in equal time correlation functions. However, when used to compute non-equal time correlation functions that are expected to show spectral properties, the mean-field theory fails to capture the bound state.

\section*{Acknowledgements}
This work was supported by the EPSRC under grant EP/S020527/1. We
are grateful to A. Das for drawing our attention to
Ref.~\cite{Haldar_2021} and helpful discussions.

\begin{appendix}

    \section{Reality of certain mean-fields}
  \label{app:reality}
When evaluating our self-consistent mean-fields we observe that some
of them are real. In this appendix we explain why this is the case, beginning with a clarification of the site parity $\sigma^\alpha_j \mapsto \sigma^\alpha_{-j}$. This does not act on the fermions as $c_j\mapsto c_{-j}$ due to the presence of the Jordan-Wigner string. The simplest way to deduce the effect of site parity in the fermion basis is to look at the action of site parity on fermion bilinears, which can be simply related to spin operators without semi-infinite strings. 
In particular, we consider the following spin bilinears of definite parity
    \begin{align}
        A =& \sigma^x_i \sigma^x_{i+1} \ , \nn         
        B =& \sigma^y_i \sigma^y_{i+1} \ , \nn
        C^\pm =& \sigma^x_i \sigma^x_{i+1} \pm \sigma^y_i \sigma^y_{i+1} \ . 
    \end{align}
We then note that the fermionic bilinears can be decomposed in terms of these via
\begin{align}
    c_i^\dag c_j =& \frac{1}{4}(A+B-iC^-) \ , \nn 
    c_j^\dag c_i =& \frac{1}{4}(A+B+iC^-) \ , \nn 
    c_i^\dag c_j^\dag =& \frac{1}{4}(A-B-iC^+) \ , \nn 
    c_i c_j =& \frac{1}{4}(-A+B-iC^+) \ .
\end{align}
We thus see that the action of site parity on the bilinears is to exchange $c_i^\dag c_j$ with $c_j^\dag c_i$ and therefore $t_{ij}=\langle c_i^\dag c_j \rangle = t_{ji} \in \R$ as stated in the main text.

Additionally, the ANNNI Hamiltonian satisfies $H=H^*=H^T$ in both the spin and
fermion bases. In particular, in the fermion basis $c_i c_j$ is also
real. By the spectral theorem for real symmetric matrices we then know
that the eigenvectors of $H$ are real in the same basis and so 
\begin{equation}
    \langle c_i c_j \rangle_\beta = \frac{1}{Z(\beta)}\sum_n \langle E_n | c_i c_j | E_n \rangle e^{-\beta E_n} \in \R    
\end{equation}
is manifestly real in equilibrium. However, after the quench the
corresponding time-evolved quantity becomes 
\begin{equation}
    \langle c_i c_j \rangle_t = \frac{1}{Z(\beta)}\sum_{n,n',m'}e^{-\beta E^0_{n}}\langle E_n| E_{m'} \rangle \langle E_{m'} |c_i c_j | E_{n'} \rangle \langle E_{n'} | E_{n} \rangle e^{-it(E_{m'}-E_{n'})} \ , 
\end{equation}
where $E^0_{n}$ are the pre-quench energies and $E_{m'}$ the
post-quench energies. Even if the post-quench Hamiltonian is also real
and thus the post-quench energy eigenstates $|E_{n'}\rangle$ real, the
phase factors will cause it to be generically complex. However, at
very late times we would expect that the system would come back to
equilibrium via these factors dephasing and so the correlation
function should become real again at late times. Since $t_n$ are all
real due to the site parity $\Z_2$ this implies that all effective
couplings are real in equilibrium, and out of equilibrium the only
complex one will be $\Delta_{\rm Eff}^{(1)}(t)$.  

\section{Linear response}
\label{App:LinearResponse}

In this appendix we summarize how to derive Kubo linear response
relations after a quantum quench that occurs at time
$t=0$, see e.g. Ref.~\cite{Rossini2014Quantum}. The Hamiltonian is of the form   
\begin{equation}
\label{eq:Htimeedp}
    H(t) = \theta(-t)H_i+\theta(t)H_f + f(t) V \ , 
\end{equation}
where $\theta(t)$ is the Heaviside step function. If $f=0$ this corresponds to a quench at $t=0$. The linear response regime is when $f(t)\ll 1$ and for this to be genuinely non-equilibrium we require $f(t)$ to have support in the time period before the system thermalizes after the quench.

We work in an interaction picture such that $H=H_0+f(t)V$, where $H_0$
is generally not free. The interaction picture states $|\psi(t)\rangle_I$ are defined by
\be
|\psi(t)\rangle_I =  e^{iH_0t}U(t,t_0)|\psi(t_0)\rangle \ ,
\ee
where $U(t,t_0)$ is the full time-evolution operator associated with $H(t)$, $|\psi(t_0)\rangle$ is the Schrödinger picture state at $t_0$ and $H_0$ is considered time independent by requiring, according to (\ref{eq:Htimeedp}), that $t_0\geq0$.  
Consistently, in the interaction picture the general operator $\mathcal{O}$ evolves in time as 
\be
\mathcal{O}_I(t) = e^{iH_0t}\mathcal{O}e^{-iH_0t}  \ .
\ee
The time-evolution according to $H(t)$ of the expectation value of $\mathcal{O}$ in the state defined at $t_0$ by $\rho(t_0) = |\psi(t_0)\rangle \langle \psi(t_0) |$ can be expressed in the interaction picture as
\be
\label{eq:linrespoofeq}
\Tr\left(\rho(t) \, {\cal O}\right) = \Tr\left(\rho_I(t) \, {\cal O}_I(t)\right) \approx \Tr\left(\rho(t_0)\,\mathcal{O}_I(t)\right) - i \int_{t_0}^t f(t')   \chi(t,t') {\rm d}t'\ ,
\ee
where the susceptibility $\chi(t,t')$ is given by
\be
\chi(t,t')\equiv \Tr\left(\rho(t_0) [{\cal O}_I (t),V_I(t') ] \right) \ .
\ee
In the last step of (\ref{eq:linrespoofeq}) we have expressed $\rho_I(t) =|\psi(t)\rangle_I {}_I \langle \psi(t)|$ by the first two terms in its power series in the small function $f(t)$. 
Eq (\ref{eq:linrespoofeq}) is the usual linear response formula except that the
time-translation invariance of the susceptibility is broken by the
quench and hence $\chi(t,t')$ does not depend only  on the
time-difference $t-t'$. 


\end{appendix}

\bibliography{references.bib}

\nolinenumbers

\end{document}